\documentclass[aps,prl,twocolumn,showpacs,floats,floatfix,letterpaper,nofootinbib,superscriptaddress,]{revtex4-1}

\usepackage{epsfig}

\begin{document}


\title{CMB Neutrino Mass Bounds and Reionization}

\author{Maria Archidiacono}
\affiliation{Physics Department and INFN, Universita' di Roma 
	``La Sapienza'', Ple.\ Aldo Moro 2, 00185, Rome, Italy}
\affiliation{Center for Cosmology, Department of Physics \& Astronomy, 
	University of California, Irvine, California 92697, USA}
\author{Asantha~Cooray}
\affiliation{Center for Cosmology, Department of Physics \& Astronomy, 
	University of California, Irvine, California 92697, USA}
\author{Alessandro~Melchiorri}
\affiliation{Physics Department and INFN, Universita' di Roma 
	``La Sapienza'', Ple.\ Aldo Moro 2, 00185, Rome, Italy}
\author{Stefania Pandolfi}
\affiliation{ICRA and INFN, Universita' di Roma 
	``La Sapienza'', Ple.\ Aldo Moro 2, 00185, Rome, Italy}

\begin{abstract}
Current cosmic microwave background (CMB) bounds on the sum of the neutrino masses 
assume a sudden reionization scenario described by a single
parameter that determines the onset of reionization.  We investigate the bounds
on the neutrino mass in a more general reionization scenario based on a principal component
approach. We found the constraint on the sum of the neutrino masses from CMB  data 
can be relaxed by a $\sim$40\% in a generalized
reionization scenario. Moreover, the amplitude of the r.m.s. mass fluctuations $\sigma_8$
is also considerably lower providing a better consistency with the low amplitude of the Sunyaev-Zel'dovich
signal recently found by the South Pole Telescope.
\end{abstract}


\maketitle



\section{Introduction}

The high precision measurements of Cosmic Microwave Background (hereafter, CMB) 
anisotropies made by the Wilkinson Microwave Anisotropy Probe (WMAP) satellite have provided 
not only a wonderful confirmation of the standard model of
cosmological structure formation but also relevant information on key parameters 
in particle physics. One example is the sum of the neutrino masses and the neutrino mass hierarchy. 

The most recent data release  from WMAP after seven years of observations presented a bound on the total 
neutrino mass of $\Sigma m_{\nu} <1.3$eV at the 95\% c.l. \cite{wmap7}. This bound is
approximately a factor five better than the current laboratory experimental  upper limit inferred from
a combination of beta-decay experiments and neutrino oscillation data (see e.g. \cite{fogli}).
The CMB bound on neutrino masses is also considered the most conservative limit
from cosmology. Indeed, including information from galaxy clustering and luminosity 
distance data the constraint can be further improved to $\Sigma m_{\nu} <0.55$eV at 95\% c.l. \cite{wmap7},
while a limit of $\Sigma m_{\nu} < 0.28$eV at $95 \%$ c.l. can be obtained 
by including redshift-dependent halo bias-mass relations \cite{fdebe}. 

It is however important to be aware of the theoretical modelling behind the  constraint based on cosmological 
measurements. A model of structure formation based on dark matter, adiabatic primordial fluctuations
and dark energy is assumed and the removal of one of these assumptions
can in principle affect the CMB limit. For example, the inclusion of 
isocurvature perturbations \cite{zunckel}, dark energy \cite{wmap7} or modified gravity \cite{tom}
can all relax the CMB upper limit on the neutrino masses.

In this {\it Brief Report} we investigate another possible theoretical caveat that could
affect the CMB bound on the sum of the neutrino masses, i.e. the modelling of the reionization
epoch. It is often assumed in the current cosmological data analysis that reionization
is a sudden event at redshift $z=z_{re}$, i.e. this process is usually described by a single
parameter with the free electron fraction $x_e$ increasing from $\sim 10^{-4}$ up to $1$ for
redshifts $z < z_{re}$ ($\sim 1.08$ for $z<3$ when taking into account Helium reionization).
While this scenario can properly describe several reionization scenarios, it can't obviously describe
more complex reionization scenarios as for example double or not-monotone reionization.
Given our current ignorance about the thermal history of the universe at redshifts
$z \ge 6$ it is important to consider all the possible reionization scenarios allowed by data when
 deriving the most conservative constraint on a cosmological parameter such as the sum of the neutrino masses.

Here, we indeed assume a more general reionization model following the 
principal components method suggested by Mortonson and Hu \cite{mortonson} and
we derive constraints on the neutrino mass in this different theoretical framework.
It has been shown recently \cite{stefania} that a general reionization scenario
can drastically alter the conclusions on inflationary parameters as the
scalar spectral index $n$, putting its value in better agreement
with the expectations of a Harrison-Zel'dovich \cite{hz}, $n=1$, spectrum. So it
is definetely timely to investigate what kind of impact a general reionization
scenario can have on the current CMB neutrino bound.
The paper is organized as follows: in the next Section we briefly describe the
reionization parametrization assumed and the decomposition in principal components.
We also describe the data analysis method. In Section III we present the results
of our analysis and in Section IV we discuss our conclusions.

\section{Analysis Method}

We adopt the method, developed in Ref.\ \cite{mortonson}, based on principal components
that provide a complete basis  for describing the effects of reionization on
large-scale $E$-mode polarization.  Following Ref.\ \cite{mortonson}, one can
parametrize the reionization history  as a free function of redshift by
decomposing $x_e(z)$ into its principal components:
\begin{equation}
x_e(z)=x_e^f(z)+\sum_{\mu}m_{\mu}S_{\mu}(z),
\end{equation}
where the principal components, $S_{\mu}(z)$,  are the eigenfunctions of the
Fisher matrix that describes  the dependence of the polarization spectra on
$x_e(z)$ (again, see Ref.\ \cite{mortonson}), $m_{\mu}$ are the amplitudes of
the  principal components for a particular reionization history, and  $x_e^f(z)$
is the WMAP fiducial model at which the Fisher matrix is computed and from which
the principal components are obtained. In what follows we use the publicly
available $S_{\mu}(z)$ functions and varied the amplitudes $m_{\mu}$ for
$\mu=1,...,5$ for the first five eigenfunctions. The eigenfunctions are 
computed in $95$ bins from redshift $z_{min}=6$ to redshift $z_{max}=30$
with $x_e(z)=1.08$ for $z<3$, $x_e(z)=1.0$ for
 $3 \le z<6$ and $x_e(z)=10^{-4}$ for $z\ge30$..
Hereafter we refer to this
method as the MH (Mortonson-Hu) case.

We have then modified the Boltzmann CAMB code \cite{camb} incorporating the 
generalized MH reionization scenario as in \cite{mortonson} 
and extracted cosmological parameters from
current data using a Monte Carlo Markov Chain (MCMC) analysis based on the
publicly available MCMC package \texttt{cosmomc} \cite{Lewis:2002ah}.

We consider here a flat $\Lambda$CDM universe described by a set of cosmological
parameters
\begin{equation}
 \label{parameter}
      \{\omega_b,\omega_c,\omega_{\nu},
      \Theta_s, n, \log[10^{10}A_{s}] \},
\end{equation}
where $\omega_b\equiv\Omega_bh^{2}$ and $\omega_c\equiv\Omega_ch^{2}$ are the
physical baryon and cold dark matter densities  relative to the critical
density, $\omega_{\nu}$ is the physical energy density in massive neutrinos,
$\Theta_{s}$ is the ratio of the sound horizon to the angular diameter
distance at decoupling, $A_{s}$ is the amplitude of the primordial spectrum, and
$n$ is the scalar spectral index. We assume $3$ degenerate, massive neutrinos
with the same mass:
\begin{equation}
m_{\nu}={30.8 eV}\times\omega_{\nu}
\end{equation}
In what follows we will use as standard parameter the value $\Sigma m_{\nu}=3m_{\nu}$.

The extra parameters needed to describe the reionization are the five amplitudes
of the eigenfunctions for the MH case
 and one single common parameter, the optical depth $\tau$,
for the sudden reionization case. 

Our basic data set is the seven--year WMAP data \cite{wmap7}  (temperature and
polarization) with the routine for computing the likelihood supplied by the WMAP
team. 

\section{Results}

\begin{table}[!htb]\footnotesize
\begin{center}
\begin{tabular}{|c|c|c|}
\hline
Parameter & WMAP7 & WMAP7 \\
& (Sudden Reionization) & (MH Reionization) \\
\hline
$\Omega_bh^2$ & ${0.0221}_{-0.0012}^{+0.0012}$ &${0.0226}_{-0.0014}^{+0.0015}$ \\
$\Omega_ch^2$ & ${0.117}_{-0.013}^{+0.013}$ &${0.115}_{-0.017}^{+0.017}$ \\
$\theta_s$ & ${1.038}_{-0.005}^{+0.005}$ &${1.039}_{-0.005}^{+0.006}$\\
$n$ & ${0.955}_{-0.033}^{+0.032}$&${0.975}_{-0.0434}^{+0.0448}$\\
$H_0$ & ${65.7}_{-8.2}^{+7.6}$ &${66.0}_{-9.0}^{+10.2}$\\
$\Omega_\Lambda$ & ${0.674}_{-0.134}^{+0.091}$ &${0.675}_{-0.148}^{+0.112}$\\
$\Sigma m_\nu$ & $<{1.15}$eV &$<{1.66}$eV\\
\hline
\end{tabular}
\caption{$95 \%$ c.l. errors on cosmological parameters in the case of sudden reionization and
MH reionization. The upper limit on the neutrino mass is relaxed by $\sim 43 \%$.}
\label{tab:results}
\end{center}
\end{table}

In Table~I we compare the constraints on several cosmological parameters
in the case of standard or MH reionization scenario.
As we can see from the table, the CMB constraint  on the neutrino mass is weakened
by $\sim 40 \%$ when a more general reionization scenario is considered.
This is not simply due to an increase in the parameter space but also due to
degeneracies present between the cosmological parameters. Considering the MH reionization
scenario renders values of the spectral index $n$ in better agreement with the
Harrison-Zel'dovich $n=1$ value (see \cite{stefania}). This changes the relative amplitude
of the peaks in the CMB angular spectrum 
and makes models with higher neutrino mass more consistent with the WMAP data.
Introducing a neutrino mass has indeed the effect of decreasing the gravitational
potential at recombination, increasing the small scale CMB anisotropy\footnote{The effect
of neutrino mass on CMB lensing for the WMAP data is negligible.}. This can
be counterbalanced by decreasing the value of the spectral index $n$ as clearly shown
by the anti-correlation in the $n$-$\Sigma m_{\nu}$ plane. A general reionization
scenario brings higher values of $n$ in agreement with observations, immediately
resulting in a better compatibility of larger neutrino masses.
It is worth noticing that while in the standard reionization scenario HZ spectra
are excluded at about three standard deviations when massive neutrinos are included
in the analysis, in the MH case the $n=1$ spectra are well consistent with the data
and inside the $1 \sigma$ c.l. also with $\Sigma m_{\nu} \sim 0.5$eV. 

\begin{figure}[h!]
\begin{center}
\includegraphics[width=8cm]{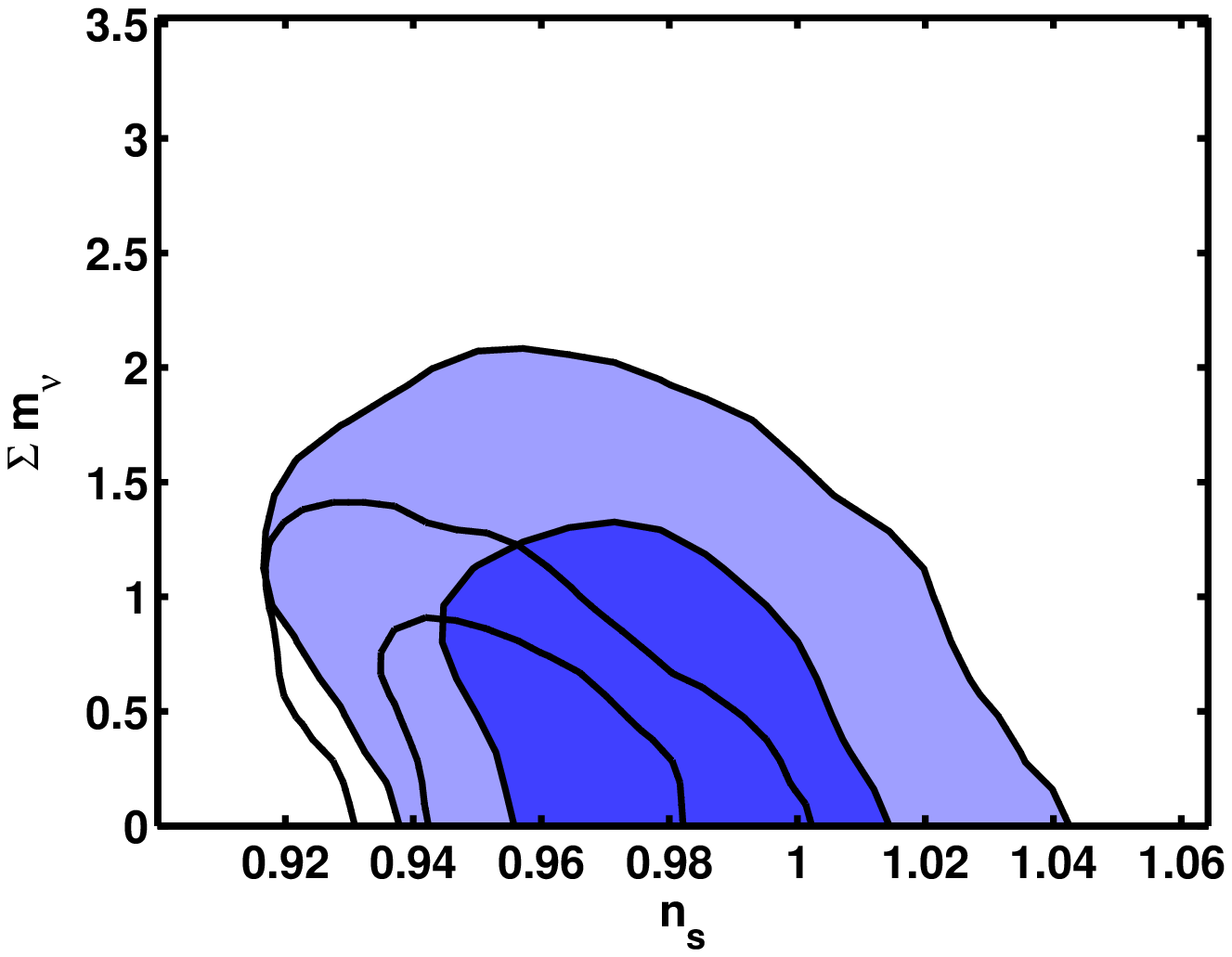}
\caption{Constraints on the $\Sigma m_{\nu}$ vs $n$ plane.
The filled contours assume MH reionization while the empty contours assume standard, 
sudden, reionization.}
\label{mnuns}
\end{center}
\end{figure}

\begin{figure}[h!]
\begin{center}
\includegraphics[width=8cm]{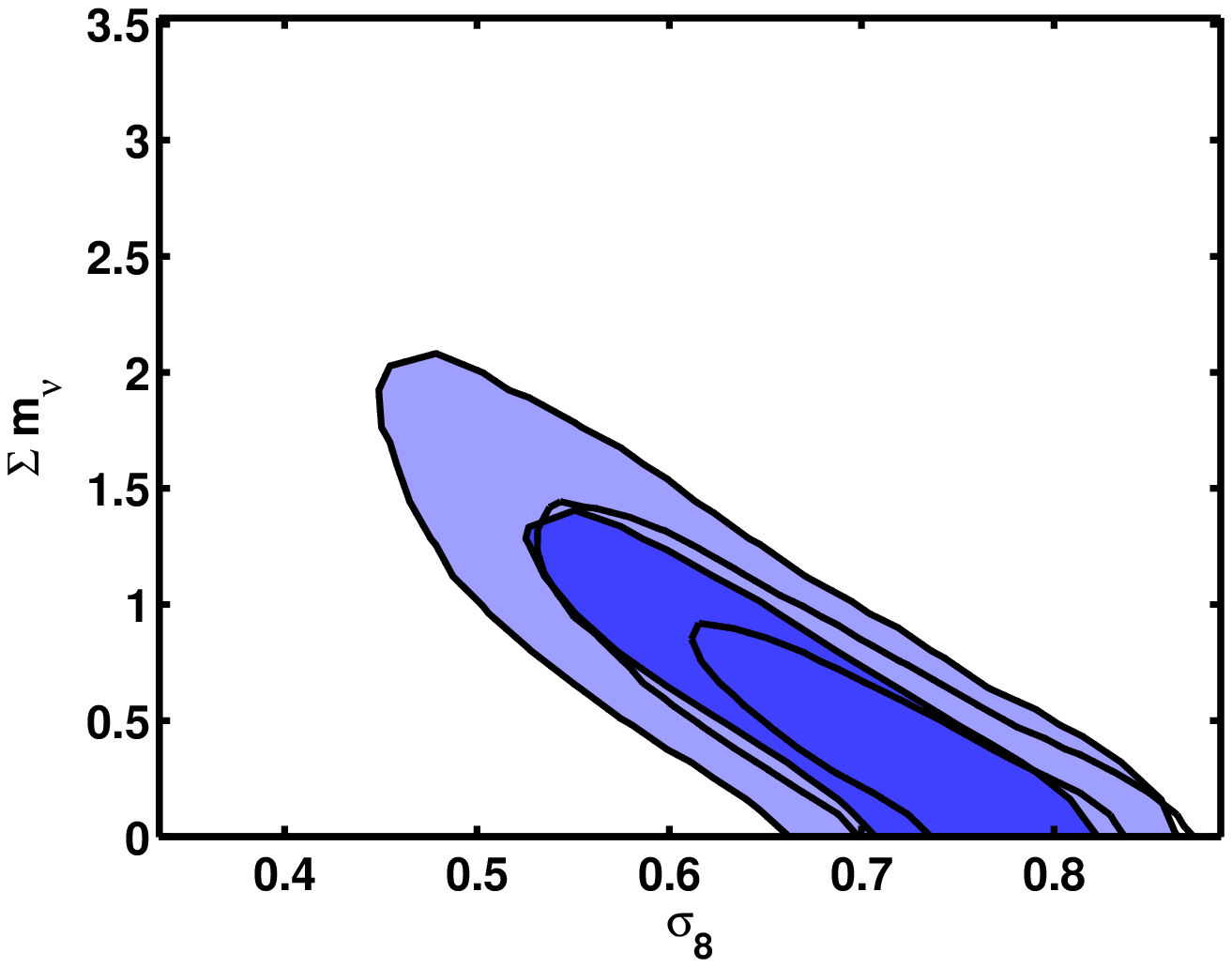}
\caption{Constraints on the $\Sigma m_{\nu}$ vs $\sigma_8$ plane.
The filled contours assume MH reionization while the empty contours assume standard, 
sudden, reionization.}
\label{mnus8}
\end{center}
\end{figure}

In Figure \ref{mnuns} we show the constraints on the $\Sigma m_{\nu}$ vs $n$ plane,
while in Figure \ref{mnus8} we show the constraints on the $\Sigma m_{\nu}$ vs $\sigma_8$ plane.
The filled contours assume MH reionization while the empty contours assume standard, 
sudden, reionization.
As we can see, MH reionization allows for values of the spectral index $n$ closer
to $1$ (as already pointed out in \cite{stefania}), for a larger neutrino mass and
for a lower $\sigma_8$ amplitude. 
It is interesting to note that a neutrino mass can in principle accomodate lower
values of $\sigma_8$ with CMB data. When MH reionization is assumed even lower
values of $\sigma_8$ are consistent with CMB data. A low value of $\sigma_8 \sim 0.77$
is preferred by the recent detection of diffuse Sunyaev-Zel'dovich effect by
the South Pole Telescope \cite{spt} experiment. 
 
\begin{figure}[h!]
\begin{center}
\includegraphics[width=8cm]{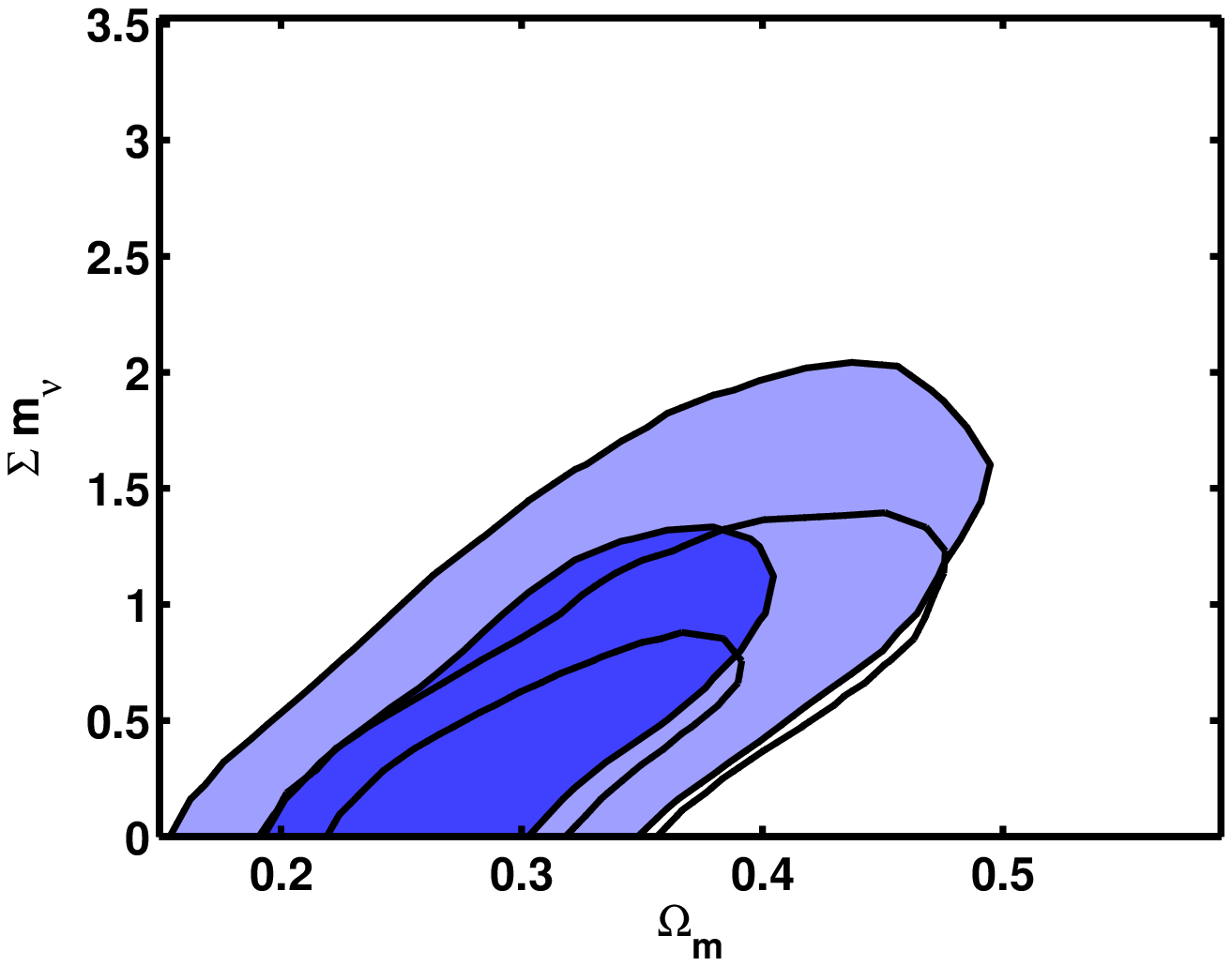}
\caption{Constraints on the $\Sigma m_{\nu}$ vs $\Omega_m$ plane.
The filled contours assume MH reionization while the empty contours assume standard, 
sudden, reionization.}
\label{mnuom}
\end{center}
\end{figure}

\begin{figure}[h!]
\begin{center}
\includegraphics[width=8cm]{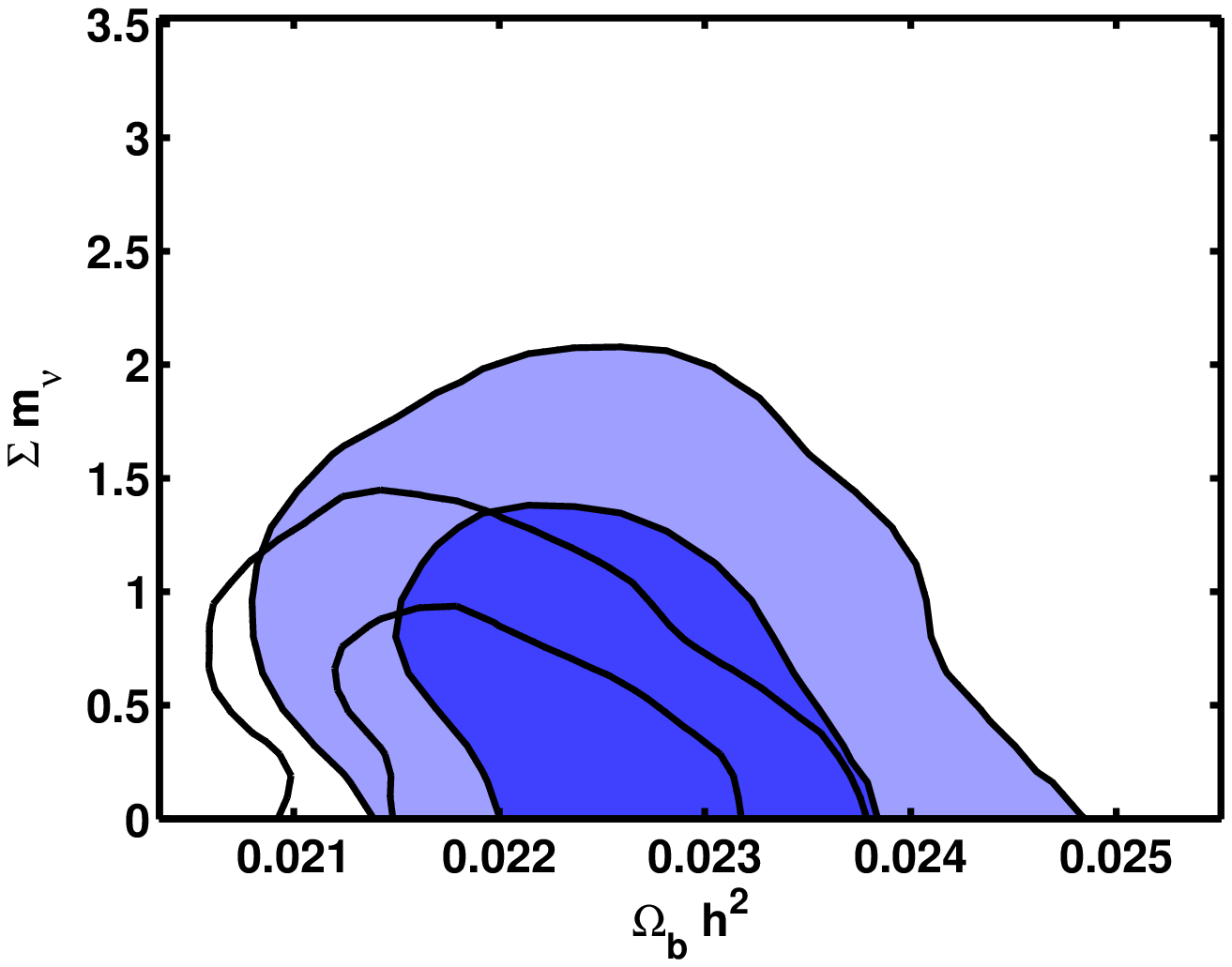}
\caption{Constraints on the $\Sigma m_{\nu}$ vs $\Omega_bh^2$ plane.
The filled contours assume MH reionization while the empty contours assume standard, 
sudden, reionization.}
\label{mnuob}
\end{center}
\end{figure}

Moreover, correlations exist with the matter density $\Omega_m$, as we show in Figure \ref{mnuom}
and (even if less pronounced) with the baryon physical density $\Omega_bh^2$ as we show in Figure \ref{mnuob}.


\section{Conclusions}

In conclusion, the details of the reionization processes in the late universe
are not very well known. In the absence of a precise, full redshift evolution
of the ionization fraction during the reionization period, a simple
parametrization, with a single parameter $z_r$, has become the standard
reionization scheme in numerical analyses. However, more general reionization
scenarios are certainly plausible and their impact on the cosmological
constraints should be carefully explored. 

In this {\it Brief Report} we have investigated the stability of the CMB constraints on 
neutrino masses in generalized reionization scenarios. We have found that a more general
treatment of reionization could potentially weaken  the current CMB upper limit on $\Sigma m_{\nu}$
by $\sim 40 \%$. 
Cosmological information from BAO for example can be added in order to reduce the uncertainty
on the neutrino mass. However the lack of knowledge on dark energy and the assumption made with regards to 
the  equation of state could again affect the neutrino mass limit with large-scale structure data.
Future data expected from the Planck \cite{planck} satellite on large angular scale CMB polarization
will help in clarifying the thermal history of the Universe and in ruling out exotic
reionization scenarios that are still in agreement with present-day observations with WMAP.



\end{document}